\documentclass[sigconf]{acmart}
\renewcommand\footnotetextcopyrightpermission[1]{} 
\usepackage{booktabs}
\usepackage{makecell}
\usepackage{multirow}
\usepackage{colortbl}
\usepackage{amsthm}
\usepackage{longtable}
\usepackage[noend]{algpseudocode}
\usepackage{algorithmicx,algorithm}
\newtheorem{myDef}{Definition}

\definecolor{mygray}{gray}{.9}
\definecolor{mypink}{rgb}{.99,.91,.95}
\definecolor{mycyan}{cmyk}{.3,0,0,0}
\def\BibTeX{{\rm B\kern-.05em{\sc i\kern-.025em b}\kern-.08emT\kern-.1667em\lower.7ex\hbox{E}\kern-.125emX}}

\begin{document}

%
\title{Bidirectional RNN-based Few-shot Training for Detecting Multi-stage Attack}

%
\author{Di Zhao}
\email{17125268@bjtu.edu.cn}
\orcid{1234-5678-9012}
\affiliation{%
  \institution{Beijing Jiaotong University}
  \streetaddress{No.3, Shangyuan Village}
  \city{Haidian}
  \state{Beijing}
  \country{China}
  \postcode{100044}
}
\author{Jiqiang Liu}
\email{jqliu@bjtu.edu.cn}
\orcid{1234-5678-9012}
\affiliation{%
  \institution{Beijing Jiaotong University}
  \streetaddress{No.3, Shangyuan Village}
  \city{Haidian}
  \state{Beijing}
  \country{China}
  \postcode{100044}
}
\author{Jialin Wang}
\email{17120417@bjtu.edu.cn}
\orcid{1234-5678-9012}
\affiliation{%
  \institution{Beijing Jiaotong University}
  \streetaddress{No.3, Shangyuan Village}
  \city{Haidian}
  \state{Beijing}
  \country{China}
  \postcode{100044}
}
\author{Wenjia Niu*}
\email{niuwj@bjtu.edu.cn}
\orcid{1234-5678-9012}
\thanks{*Corresponding author: Wenjia Niu, Email: niuwj@bjtu.edu.cn}
\affiliation{%
  \institution{Beijing Jiaotong University}
  \streetaddress{No.3, Shangyuan Village}
  \city{Haidian}
  \state{Beijing}
  \country{China}
  \postcode{100044}
}
\author{Endong Tong*}
\email{edtong@bjtu.edu.cn}
\thanks{*Corresponding author: Endong Tong, Email: edtong@bjtu.edu.cn}
\orcid{1234-5678-9012}
\affiliation{%
  \institution{Beijing Jiaotong University}
  \streetaddress{No.3, Shangyuan Village}
  \city{Haidian}
  \state{Beijing}
  \country{China}
  \postcode{100044}
}

\author{Tong Chen}
\email{18112044@bjtu.edu.cn}
\orcid{1234-5678-9012}
\affiliation{%
  \institution{Beijing Jiaotong University}
  \streetaddress{No.3, Shangyuan Village}
  \city{Haidian}
  \state{Beijing}
  \country{China}
  \postcode{100044}
}

\author{Gang Li}
\email{Gang.li@deakin.edu.au}
\orcid{1234-5678-9012}
\affiliation{%
  \institution{Deakin University}
  \state{Victoria}
  \country{Australia}
}
%
\begin{abstract}
"Feint Attack", as a new type of APT attack, has become the focus of attention. It adopts a multi-stage attacks mode which can be concluded as a combination of virtual attacks and real attacks. Under the cover of virtual attacks, real attacks can achieve the real purpose of the attacker, as a result, it often caused huge losses inadvertently. However, to our knowledge, all previous works use common methods such as Causal-Correlation or Cased-based to detect outdated multi-stage attacks. Few attentions have been paid to detect the "Feint Attack", because the difficulty of detection lies in the diversification of the concept of "Feint Attack" and the lack of professional datasets, many detection methods ignore the semantic relationship in the attack. Aiming at the existing challenge, this paper explores a new method to solve the problem. In the attack scenario, the fuzzy clustering method based on attribute similarity is used to mine multi-stage attack chains. Then we use a few-shot deep learning algorithm (SMOTE\&CNN-SVM) and bidirectional Recurrent Neural Network model (Bi-RNN) to obtain the "Feint Attack" chains. "Feint Attack" is simulated by the real attack inserted in the normal causal attack chain, and the addition of the real attack destroys the causal relationship of the original attack chain. So we used Bi-RNN coding to obtain the hidden feature of "Feint Attack" chain. In the end, our method achieved the goal to detect the "Feint Attack" accurately by using the LLDoS1.0 and LLDoS2.0 of DARPA2000 and CICIDS2017 of Canadian Institute for Cybersecurity.
\end{abstract}

%
%


%

%

%
\maketitle

\section{Introduction}
Under the background of the rapid development of global network informationization, the hidden, pervasive and targeted Advanced Persistent Threat (APT) poses a growing threat to various high-level information security systems ~\cite{Li2016The}. APT attacks are increasing which target to national and enterprise network information systems and data security face severe challenges. In March 2011, at the 6th International Conference on Information Warfare and Security (ICIW), three security researchers at Lockheed Martin proposed an Intrusion Kill Chain (IKC) ~\cite{Bhatt2014Towards}. From the perspective of intrusion detection, they decompose the attack process into seven steps: reconnaissance, weaponization, delivery, exploitation, installation, command and control (C2), and actions on objectives. This model redefines the kill chain in the military field to cyberspace security, providing us with new ideas for solving APT attacks ~\cite{Yadav2015Technical}~\cite{Hahn2015A}.

However, at the end of 2017, Trend Micro pointed out that there has been a new type of APT attack named "Feint Attack" ~\cite{6636288}. It not only uses the same attack, but also makes full use of two separate malware attacks. One attack (Virtual attack) is responsible for distracting and masking the malicious activity of another attack (Real attack) to provide a way to further infect or steal data and intellectual property. \emph{Enterprise IT Security Risk Survey Report} pointed out that the above-mentioned virtual attacks are often distributed denial of service (DDoS) attacks ~\cite{Bhuyan2014Information}. Through analysis of security experts, these DDoS attacks are only "smoke bombs" that attacks use to cover their real attacks. Some enterprises that have suffered DDoS attacks finding that DDoS attacks are only part of the overall network attack, accounting for only 29\% of the total attack time. When a DDoS attack occurs, the enterprise's security department must try to quickly restore normal access services because the normal external access of the enterprise is denied or interrupted. Therefore, during the DDoS attack, security departments are often required to go all out to solve the DDoS attack problem, and then the attacker "make a feint to the east but attack in the west" and cannot take into account the other intrusion. After the "Feint Attack", 25\% of companies will lose important data at the same time. As it turns out, in order to improve the efficiency of attackers, an attacker often launches a variety of other forms of attack when launching a DDoS attack. Therefore, once a company is found to be attacked by DDoS, it must understand the full threat situation and be ready to handle multiple types of network attacks, otherwise it is likely to suffer greater losses. The Trend Micro report predicted that such attacks will become more common in 2018.

The "Feint Attack" mode has received extensive attention in the field of cyberspace security. However, in the face of special attacks, how to carry out related detection and defense work is still a problem. The detection of multi-stage attack mode at home and abroad is currently in the key research stage. This paper mainly focuses on the special attack mode of "Feint Attack", and proposes a detection model based on fuzzy clustering in alert correlation and Bi-RNN algorithm. The main contributions are as follows:

\begin{itemize}
\item[(1)] Replaying the traffic packet of the LLDoS 1.0 and LLDoS2.0 of DARPA2000 intrusion detection attack scenario ~\cite{DARPA} and the traffic packet (.pcap) of Intrusion Detection Evaluation Dataset (CICIDS2017) ~\cite{CICIDS} ~\cite{Sharafaldin2018Toward} through snort, generating the raw alert data, further based on the five-tuple (AttackType, S\_IP, D\_IP, S\_Port, D\_Port) performs alert aggregation. The main purpose is to reduce the duplicate alert data of the same attack event, and use the fuzzy clustering based on attribute similarity to process the raw alert after aggregation. Multi-stage attack chains are mined in the attack scenario to form a multi-stage attack mode comparison library.
\item[(2)] We improved the traditional deep learning algorithm, CICIDS2017 dataset in our experiment is preprocessed by the imbalanced learning strategy. We used the deep convolutional neural network to learn the new feature representation of the dataset. Then the few-shot learning is performed by the hierarchical SVM classifier. The classification result was defined and divided the virtual attacks and real attacks with the confidence level. Finally, we constructed the dataset of virtual attacks and real attacks, which is the basic element library of the "Feint Attack" chain.
\item[(3)] Using the multi-stage attack and element attack event library obtained in the first and second stages, our method of attack chain recovery technology based on Bi-RNN. According to the method, "Feint Attack" is simulated by the real attack inserted in the normal causal attack chain, and the addition of the real attack destroys the causal relationship of the original attack chain. The hidden feature is obtained by Bi-RNN coding. Further we classified the two types of trainable samples. Finally, our work achieved the purpose of detecting the "Feint Attack" accurately.
\end{itemize}
The structure of this paper is organized as follows. Section 2 will discuss the related work in this field. Section 3 will present "Feint Attack" chains construction and detection methods through "Feint Attack" chains model. Section 4 gives experimental and results. Finally, conclusion is showed in Section 5.
\section{Related Work}
In this section, we review the related work about approaches that detect multi-stage attacks using IDS alerts. Generally, the approaches may be classified into two categories, namely causal correlation analysis and cluster correlation analysis.

\textbf{Causal Correlation Analysis}

The causal alert correlation method associates the alert information according to the causal dependence between the attacks. If the result of one attack behavior creates a precondition for another attack behavior, it is considered that there is a causal dependence between the two attack behaviors, and the causal relationship is utilized. Nguyen et al.~\cite{Nguyen2017Multi} conducted an empirical game analysis of the multi-stage interaction between the attacker and the defender to obtain a heuristic strategy under the Bayesian attack graph model. Haas et al.~\cite{GAC} proposed a graph-based alert association (GAC) algorithm to isolate attacks and identify attack scenarios, and assemble multi-stage attacks from a large set of alerts. Pei et al.~\cite{Pei2016HERCULE} proposed a method which model multi-stage intrusion analysis as a community discovery problem analysis system, and discovers any "attack communities" embedded within the graphs. A novel method based on the Hidden Markov Model is proposed to predict multi-stage attacks using IDS alerts by Holgado et al.~\cite{Holgado2017Real} They consider the hidden states as similar phases of a particular type of attack. Katipally et al.~\cite{Katipally2011Attacker} use data mining to process alarms and input the processed data into the hidden Markov model (HMM), ultimately achieving the purpose of analyzing and predicting the behavior of the attacker.

\textbf{Cluster Correlation Analysis}

The clustering alert correlation method associates alert information with some identical or similar features, that is, clustering by the similarity between alert attribute values, such as the same destination address, the same attack source, attack means, etc. Ahmadianramaki et al.~\cite{Causal} proposed a three-layer processing framework that uses causal knowledge to correlate alerts, automatically extracts causal relationships between alerts, builds the attack scenario using Bayesian networks. And further predict the most likely next attack behavior. Barzegar et al.~\cite{Barzegar2018Attack} proposed approach reconstructs attack scenarios by reasoning based on the evidences in the alert stream. The main idea of the proposed approach is to identify the causal relation between alerts using their similarity. Alvarenga et al.~\cite{Alvarenga2017Process} approach applies process mining techniques on alerts to extract information regarding the attackers behavior and the multi-stage attack strategies they adopted. The strategies are presented to the network administrator in friendly high-level visual models. Large and visually complex models that are difficult to understand are clustered into smaller, simpler and intuitive models using hierarchical clustering techniques.

\textbf{Bidirectional RNN}

RNN is used in the field of natural language processing and its main purpose is to process and predict sequence data ~\cite{Kim2016Character}. The neural network memorizes the previous information, stores it in the internal state of the network, and applies it to the calculation of the current output, that is, the nodes between the hidden layers are no longer connected but connected, and the hidden layer The input contains not only the output of the input layer but also the output of the hidden layer at the previous moment. In the classical cyclic neural network, the state of the transmission is one-way from the back. The bidirectional RNN ~\cite{Schuster1997Bidirectional} can memorize and encode the context information, and the subject structure is the combination of two unidirectional RNNs. At each time $t$, the input is provided to both RNNs in opposite directions, and the output is determined by the two unidirectional RNNs (which can be spliced or summed, etc.). In this paper, the Bi-RNN algorithm is introduced into the coding part of the multi-stage attack sequence. In the multi-stage attack scenario, an attack chain can be analogized into a sequence. Each atomic attack is equivalent to one word. By encoding the attack sequence through Bi-RNN, the causal association of the attack sequence can be preserved to the greatest extent, and achieve the goal of reducing the dimension.

The multi-stage attack detection based on causal correlation requires a large amount of expert knowledge to support, and the acquisition of expert knowledge is very difficult, and can not discover new attack behavior. In this paper, the fuzzy clustering method based on attribute similarity is used to mine the multi-stage attack mode. The previous work of detection of the multi-stage attack chain does not consider the special type of "Feint Attack" chain, and the length of the constructed attack chain is too long, which makes it difficult to retain its inherent causal relationship in further analysis and pre-processing. Therefore, our research is based on previous work. It mainly achieve the goal of defining and dividing virtual attacks and real attacks, builds the attack chain based on causal correlation and Bi-RNN model, further obtains the trainable attack sample set, and finally obtains the attack chain detection classifier through training.

\section{"Feint Attack" Chains Construction and Detection Method}

\begin{figure}[!htbp]
  \centering
  \includegraphics[height=5.5cm,width=8.9cm]{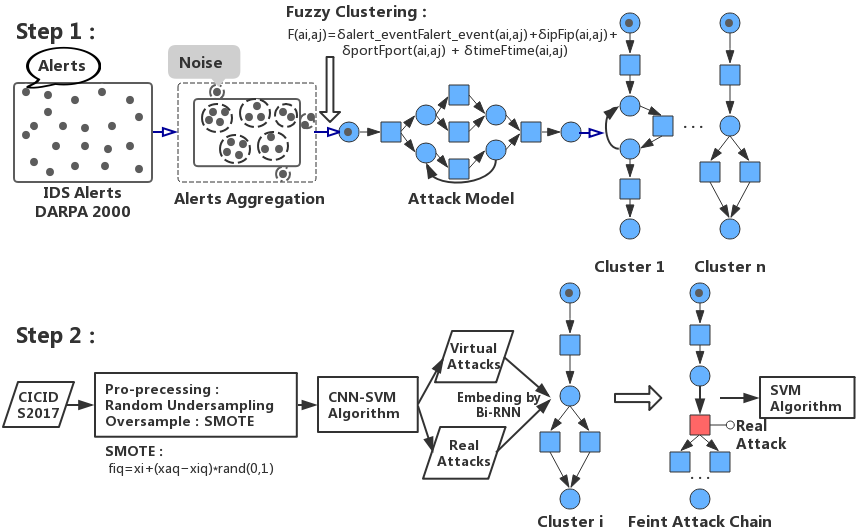}
  \caption{Framework of "Feint Attack" Chains Construction and Detection Method.}\label{fig:Framework}
\end{figure}

In order to achieve the "Feint Attack" chain detection based on the virtual attack chain and real attack chain, we try to solve the problem through proposed new detection method in this section which mainly utilizes fuzzy clustering and Bi-RNN algorithm. The input to our model is raw data stream (The packet format is .dump and .pcap) of LLDoS1.0 and LLDoS2.0 of DARPA2000 and CICIDS2017 of Canadian Institute for Cybersecurity, and the output is the result of the classifier for detect "Feint Attack". That is, whether there is a "Feint Attack" behavior in a multi-stage attack sequence. We will describe in detail the implementation of each algorithm proposed in this paper, and show how to achieve our model to construct and detect the "Feint Attack" chain. Framework of bidirectional RNN based few-shot training for detecting multi-stage attack model is shown in Figure \ref{fig:Framework}.

Using the captured real-time data packet or replaying the classical attack dataset by snort to obtain the raw alerts. The multi-stage attack mode is mined by the fuzzy clustering method based on attribute similarity, and the virtual attack and real attack are defined and divided by the few-shot deep learning model ~\cite{Few-shot}. The real attack is embedded into the attack chain by Bi-RNN coding, and the "Feint Attack" chain is constructed. Further we classified the two types of trainable samples. Finally, our work achieved the purpose of detecting the "Feint Attack" accurately.

\subsection{Alert Correlation Based on Fuzzy Clustering}

\begin{myDef}
    \textbf{IDS alert} is a kind of alert generated when attack operations occur. It shows security situation of the entire network. We represent IDS alert as $alert={a_1,a_2,...a_n}$, where $a_i$ indicates the $i_{th}$ alert and is a nine-tuple:\\
    \begin{equation*}
    \begin{aligned}\label{eps}
    a_i=(Timestamp,Protocal,S\_IP,D\_IP,S\_Port,D\_Port,\\AttackType,Classification,Priority)
    \end{aligned}
    \end{equation*}
\end{myDef}

\begin{myDef}
    \textbf{Raw alert} refers to a single attack action performed by the attacker in the network. It may be an alarm generated directly by the IDS system after the scan of the host service or the exploitation of a vulnerability of the host, without any processing.
\end{myDef}

\begin{myDef}
    \textbf{An attack sequence} is a sequence of IDS alerts that is produced by an attacking process. We represent the attack sequence as $AS=\{a_1,a_2,...a_n\}$.
\end{myDef}

\textbf{Alert Aggregation}

We found that there are many attack type, source IP, destination IP, source port and destination port with the same or similar alerts in a certain time window, which are recorded as five-tuple $(AttackType, S\_IP, D\_IP, S\_Port, D\_Port)$. According to the specific circumstances of the alert, this paper is divided into the following modes:

a) $(AttackType, S\_IP, D\_IP, S\_Port, D\_Port)$ are the same, that is the same attack event is alerted multiple times.

b) $(AttackType, S\_IP, D\_IP, S\_Port)$ is the same, an attacker scans the ports of another host and queries the services it runs.

c) $(AttackType, S\_IP)$ is the same, $D\_IP$ is on the same network segment, and an attacker scans the target network segment to query the surviving hosts.

d) The $AttackType$ is different, $S\_IP$ and $D\_IP$ are same,that is belongs to the springboard attack.

By merging multiple alerts caused by the same security event into one alert record, the alert aggregation can greatly reduce the number of raw alerts and reduce the number of alerts to be associated, which can greatly reduce the time required for alert correlation. The complexity of the resulting multi-stage attack model is greatly reduced. It is more conducive to us to explore the phenomenon of "Feint Attack".

We defined the \emph{Alert Aggregation Rate} as follows:

\begin{equation}
  Alert\ Aggregation\ Rate=\frac{Raw\ Alerts-Output\ Alerts}{Raw\ Alerts}
\end{equation}


\textbf{Attribute Similarity Calculation}

\begin{itemize}
\item \emph{\textbf{Attack Event}}
The attack events in the IDS alerts are classified based on the IKC model. From the attacker's point of view, the attacks in the subsequent stages are more complex and more purposeful, and the acquired rights are higher. In the attack event dimension, the similarity formula for $a_i$, $a_j$ belonging to an attack sequence is as follows:

\begin{equation}
F_{alert\_event}(a_i,a_j)=
\left\{
             \begin{array}{lr}
             1,\Delta\alpha=0\,or\,1   \\
             e^{-(\Delta\alpha-3/2)},\Delta\alpha>1\\
             0,else &
             \end{array}
\right.
\end{equation}

\begin{equation*}
  \Delta\alpha=\alpha(a_i.alert\_event)-\alpha(a_j.alert\_event)
\end{equation*}
Indicates the stage where the $a_i$ attack event is located, indicating the difference between the two alarms. If the alarm is 0 or 1, the greater the similarity, the smaller the similarity is. The upper limit of similarity is 1, and the minimum is 0.
\item \emph{\textbf{IP Address}}
We use the method of comparing the same number of bits of an IP address to measure the similarity of IP addresses.
\begin{equation}
  F_{IP}(a_i,a_j)=\frac{N}{32}
\end{equation}
where $N=max\{H(a_i.sIP,a_j.dIP),H(a_i.sIP,a_j.sIP),\\H(a_i.dIP,a_j.dIP)\} $

\item \emph{\textbf{Port}}
The maximum value of the port is 65535, so the port difference value can be normalized to represent.
\begin{equation}
  F_{Port}(a_i,a_j)=1-\frac{\left|p1-p2 \right|}{65535}
\end{equation}
\item \emph{\textbf{Timestamp}}
In a multi-stage attacking process, the time interval is relatively short when two attacks are in the same phase, and the time interval may be longer when two attacks occur in different phases, and when there is a long latency following the previous access. For this reason, we do not set time window for alert logs. The similarity function of the timestamp property is as follows:
\begin{equation}
\begin{split}
  \qquad\qquad F_{Time}(a_i,a_j)=e^{-\Delta t}\qquad\qquad\qquad \\
  \Delta t=a_i.time-a_j.time.
\end{split}
\end{equation}
\end{itemize}
The complete similarity is calculated using the following function:
\begin{equation}
\begin{split}
F(a_i,a_j)=&\delta_{alert\_event}F_{alert\_event}(a_i,a_j)+\delta_{ip}F_{ip}(a_i,a_j)+\\
&\delta_{port}F_{port}(a_i,a_j)+\delta_{time}F_{time}(a_i,a_j)
\end{split}
\end{equation}
$\delta$ is the weight of the attribute value.

Scan the alert sets after aggregating, analyze each alert $a_i$ in turn, and calculate the membership degree of each classified result of $a_i$. The specific calculation method is to calculate the similarity function of all alerts in $a_i$ and a cluster. The maximum degree of similarity is used as the membership of this class of $a_i$. Before calculating the similarity, first determine the attack event of the alert with the latest timestamp in the cluster and the attack event of $a_i$, whether the number of stages corresponding to the attack event of the latter is less than the number of stages corresponding to the attack event of the former is greater than or equal to -1, if it is greater than or equal to -1, we calculate the similarity of two alerts using a similarity function with multidimensional attributes. If less than -1, we calculate the membership of $a_i$ belonging to the next cluster. The largest membership degree of $a_i$ belonging to the existing clusters is $r$. When $r$ is greater than the threshold value $\lambda$, it is considered that the alerts in the clusters corresponding to the alerts $a_i$ and $r$ are triggered by the same attack process. If $r$ is less than the threshold value $\lambda$, $a_i$ is used as a new cluster, which may be the beginning of a new attack process. The specific algorithm is described as follows:

\begin{table}[!htbp]
 \setlength{\abovecaptionskip}{1cm}
 \setlength{\belowcaptionskip}{-0.cm}
  \label{tab:Algorithm2}
  \centering
  \setlength{\tabcolsep}{0.001mm}{
\begin{tabular}{p{240pt}}
   \Xhline{1.2pt}
   \textbf{Algorithm:} Fuzzy Clustering Algorithm Process\\ \hline
   \quad\textbf{Input:} $Alerts = \{a_1,a_2,...,a_n\}$, and attack sequence set \\ \qquad \qquad $ASS=\phi$\\
   \quad\textbf{Output:} Attack sequence set $ASS=\{AS_1,AS_2,...,AS_q\}$, \\ \qquad \qquad where each attack sequence $AS_i=<a_1,a_2,...,a_n>$.\\  \hline
   \quad \textcircled{1}For each raw alert $a_i$, calculate its membership to each attack sequence $AS_i$. If the attack sequence set $ASS=\{AS_1,AS_2,...,AS_q\}$ is empty, then make $AS_1=\{a_i\}$, and repeat step \textcircled{1}. If $ASS$ is not empty, then use $AS_1$ in the $ASS$ set in step \textcircled{2}.\\

   \quad \textcircled{2}Scan attack sequence $AS_i=<a_1,a_2,...,a_k>$. First determine whether the phase of $AS_i$ (the phase in which the latest timestamp in $AS_i$ occurs). If the answer is yes, go to step \textcircled{3}, and if the answer is no, then go to step \textcircled{4}.\\

   \quad \textcircled{3}Calculate the similarity between $a_i$ and each element in $AS_i$ separately using the similarity function and use the maximum value of the results as a membership degree of $a_i$ to $AS_i$. If the membership degree is greater than or equal to the preset threshold value $\lambda$, then add $a_i$ to attack sequence $AS_i=\{a_1,a_2,...,a_k,a_i\}$ and go to step \textcircled{4}.\\

   \quad \textcircled{4}Take the next $AS_i$ in $ASS$, if it exists, repeat step \textcircled{2}; if not, it means that all the attack sequences in the $ASS$ have been scanned. If the membership degree of $a_i$ to every attack sequence is less than $\lambda$, then create a new element $AS_r=\{a_i\}$ and add $AS_r$ to $ASS=\{AS_1,AS_2,...,AS_q,AS_r\}$, before going to step \textcircled{5}.\\

   \quad \textcircled{5}Repeat step \textcircled{1} to step \textcircled{4} zbove until all Alerts are analyzed.\\
   \hline
\end{tabular}}
\end{table}

\subsection{Building the Virtual-Real Lib}

\begin{myDef}
    \textbf{Virtual Attack and Real Attack:}
    A virtual attack is defined as an attack that is accurately identified by IDS or an attack classifier. A real attack is a more concealed attack (which may be a normal behavior) or a new type of attack. The IDS does not generate an alert or is judged to be a normal behavior by the classifier.
\end{myDef}

The input to this section is the CICIDS2017 dataset, which has 83 statistical features such as duration, number of packets, number of bytes, packet length and so on. The output is the classification result of the attack, which lays the foundation for the next step of dividing the virtual attacks and real attacks. It mainly studies existing attack detection algorithms and improves traditional deep learning methods for attack detection. Based on deep learning and few-shot deep learning algorithms, the raw alerts are preprocessed by unbalanced learning strategies, such as random downsampling and SMOTE oversampling techniques, combined with deep convolutional neural networks to select dataset features, and then through hierarchical SVM classifiers to build the optimal CICIDS2017 classifier. The network structure is shown in Table~\ref{tab:cnn-svm}.

\begin{table}[!htbp]
\caption{CNN Architecture for Feature Extraction and SVM for Classification}
\label{tab:cnn-svm}
\setlength{\tabcolsep}{0.05mm}{
\begin{tabular}{|c|c|c|c|}

\hline
\textbf{Index}& \textbf{Layer}& \textbf{Output Shape}& \textbf{Pad\&Stride}\\\hline
\textbf{1}& Conv2D(32 filter,size:3$\times$3)& 10$\times$10$\times$32& 1,1\\\hline
\textbf{2}& Conv2D(32 filter,size:3$\times$3)& 10$\times$10$\times$32& 1,1\\\hline
\textbf{3}& Maxpooling2D& 5$\times$5$\times$32& 2,2\\\hline
\textbf{4}& Conv2D(64 filter,size:3$\times$3)& 5$\times$5$\times$64& 1,1\\\hline
\textbf{5}& Conv2D(64 filter,size:3$\times$3)& 5$\times$5$\times$64& 1,1\\\hline
\textbf{6}& Flatten& 1600&-- \\\hline
\textbf{7}& Fully Connected& 512&-- \\\hline
\textbf{8}& SVM& 5&-- \\
\hline

\end{tabular}}
\end{table}

The results of CICIDS2017 classifier are filtered, and to enhance the confidence of the virtual attacks and real attacks, we selected the attacks that were judged to be omissive judgement in ten tests, that is, the attack events identified as normal events as real attack sample set. The correctly classified attack events are used as a set of virtual attack samples. The virtual attacks and real attacks are respectively extracted from the original dataset and store them in the file to form a virtual attack and real attack sample database, that is, the basic element library of the attack chain construction, we defined the concept of the virtual reality attack confidence as the probability virtual attack and real attack samples are judged as normal event. The lower the probability value of the virtual attack, the greater the probability that the attack is a virtual attack, and the high probability value of the real attack indicates that the attacker is more likely to be attacked by the attacker. The division of virtual attacks and real attacks is shown in Table~\ref{tab:division}. The red background in the table is real attacks, and the blue background is virtual attacks.

\begin{table}[!htbp]
\caption{Divide the virtual attack and real attack}
\label{tab:division}
\centering
\setlength{\tabcolsep}{0.00001mm}{
\begin{tabular}{|c|c|c|c|c|c|c|c|}
\hline

\multicolumn{2}{|c|}{ \multirow{2}*{\textbf{Confusion Matrix}} }& \multicolumn{5}{c|}{Predicted Category} &\multirow{2}*{\textbf{\emph{Recall}}}\\
\cline{3-7}
\multicolumn{2}{|c|}{}&Normal&Probe&DoS&U2L&R2L&\\
\hline
\multirow{5}*{Actual Category}&Normal&60352&123&103&9&6&0.996\\
\cline{2-8}
&Probe&\multicolumn{1}{>{\columncolor{mypink}}c}{387}&\multicolumn{1}{>{\columncolor{mycyan}}c}{3501}&260&0&18&0.840\\
\cline{2-8}
&DoS&\multicolumn{1}{>{\columncolor{mypink}}c}{5686}&82&\multicolumn{1}{>{\columncolor{mycyan}}c}{224081}&0&4&0.975\\
\cline{2-8}
&U2R&\multicolumn{1}{>{\columncolor{mypink}}c}{73}&13&17&\multicolumn{1}{>{\columncolor{mycyan}}c}{119}&6&0.522\\
\cline{2-8}
&R2L&\multicolumn{1}{>{\columncolor{mypink}}c}{7018}&4&6&1&\multicolumn{1}{>{\columncolor{mycyan}}c}{9160}&0.566\\
\cline{1-8}
\multicolumn{2}{|c|}{\textbf{\emph{Precision}}}&0.821&0.940&0.998&0.922&0.996&\textbf{\emph{Acc:95.6\%}}\\
\hline
\end{tabular}}
\end{table}

\textbf{Over-Sampling: SMOTE}

\begin{figure}[h]
  \centering
  \includegraphics[height=3cm,width=5.5cm]{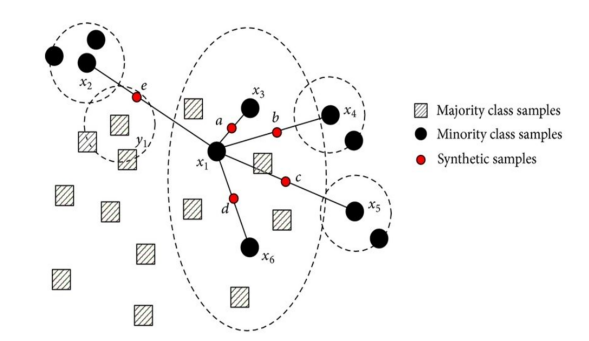}
  \caption{Schematic diagram of SMOTE algorithm.}\label{fig:smote}
\end{figure}

Less class-sample combining oversampling technique referred to as SMOTE algorithm, it is by Chawla.N ~\cite{S2015SMOTE}, who proposed one based on the traditional method of oversampling little wood class like a simple copy of the different new oversampling. In the training data $S$, $x_i$ is the minority samples. The first step to calculate the $x_i$ similar k-nearest neighbor set $P_i$. From $P_i$ random selection of a sample, it may be set to $x_a$, the difference between the $x_i$ and $x_a$ corresponding to the attribute $q$ is denoted as $diff(q)= x_{aq} - x_{iq}$. It can be concluded that the synthesis of minority class of sample $f_{iq}$ mathematical expressions is such as ~\ref{eps}.

\begin{equation}\label{eps}
f_{iq}=x_{i}+(x_{aq}-x_{iq})*rand(0,1)
\end{equation}

where rand (0,1) is expressed the random number in (0,1). Then the operation of the above process is repeated according to the beginning of the set of over sampling rate, and the synthesis of a new minority sample is added to the initial training sample to increase the number of minority samples. And the degree of imbalance will greatly reduce and get the new training samples, then there is basic balance between majority classes and minority classes in the new training data set. get the new training sample multi class sample and small sample in the number of basic balance. Finally, the new training data set are classified by the classifier and the results are obtained. The process is shown in Figure ~\ref{fig:smote}.

\subsection{"Feint Attack" Chains Construction and Detection Model}

\begin{myDef}
    \textbf{"Feint Attack" Chains:} By analyzing the various situations of "Feint Attack", it is summarized as a multi-stage attack mode of virtual attacks and real attacks.
\end{myDef}
1) The attacker hides the attack trajectory, and sometimes uses the method of "make a feint to the east but attack in the west" to perform a large number of attacks on the vital host $A$, such as DDoS attacks, generating a large number of alerts, while the real target host is $B$. The operation and maintenance personnel handle the DDoS for $A$. Attacks against $B$ when the attack does not take into account other alerts;

2) The attacker uses a highly concealed attack in some steps of the multi-stage attack sequence, or uses an advanced attack to prevent the IDS system from generating an alert to confuse the operation and maintenance personnel. The lack of some processes in the multi-stage attack process, resulting in the inability to completely restore the entire attack path (such as using DNS queries in LLDoS2.0 instead of IPsweep in LLoS1.0).
\begin{figure}[h]
  \centering
  \includegraphics[width=\linewidth]{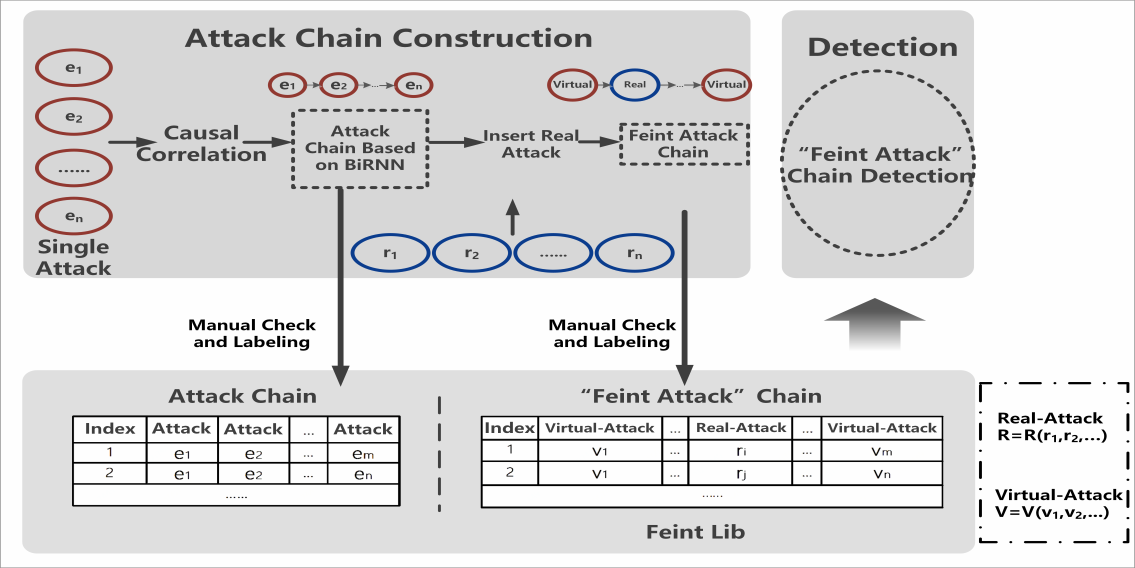}
  \caption{"Feint Attack" Chains Construction.}\label{fig:chain}
\end{figure}

Our attack chain recovery technique which based on Bi-RNN. We use the attack chain established in the first stage and the real attack in the second stage to embed the atomic attack event into the attack chain through Bi-RNN coding to construct the attack chain. The forward RNN records the information of the attack chain from the cause and the result, and the reverse RNN records the information of the attack chain from the result to the cause to ensure the maximum retention of the correlation information. The process is shown in Figure~\ref{fig:chain}.

Finally, we label the sample set of feint attack and non-feint attack chains, and further classify the attack chain samples. Based on the machine learning algorithm, a special attack detection model based on the virtual attack and real attack chain is constructed by training the feint attack and non-feint attack chain samples, and the model parameters for the data set are determined. Further, the learning model integration of specific weight enhancement is carried out by voting method to improve the accuracy of model detection. Finally, we achieve the purpose of accurately identifying the "Feint Attack".

\begin{table}[!htbp]
  \label{tab:Algorithm}
  \centering
  \setlength{\tabcolsep}{0.001mm}{
\begin{tabular}{l}
   \Xhline{1.2pt}
   \textbf{Algorithm:} "Feint Attacks" Construction and Detection Model \\ \hline
   \quad\textbf{Input:} the CICIDS2017 dataset\\
   \quad\textbf{Output:} the classifier of the "feint attack" chain\\  \hline
   \quad \textbf{Step 1: Create CNN-SVM model\_1}\\
   \qquad 1. Add 1$^{st}$ and 2$^{nd}$ convolution layers with 32 filters of 3$\times$3, \\ \qquad followed by max pooling layer of size 2$\times$2.\\
   \qquad 2. Add 3$^{rd}$ and 4$^{th}$ convolution layers with 64 filters of 3$\times$3, \\ \qquad followed by flatten layer and the out put of which is a temp \\ \qquad vector of 1600$\times$1.\\
   \qquad 3. Add fully connected to get a vector(512$\times$1), followed by a \\ \qquad H-SVMs classifier.\\
   \quad \textbf{Step 2: Build the Virtual-Real Lib}\\
   \qquad $For$ ten times of test in model\_1\\
   \qquad\qquad$If$ a attack is predicted to be $Normal$\\
   \qquad\qquad\qquad add the attack with the probability of $Normal$ to\\ \qquad\qquad\qquad \textbf{Real Lib}\\
   \qquad\qquad$Else~ if$ a attack is predicted correctly\\
   \qquad\qquad\qquad add the attack with the probability of $Normal$ to\\ \qquad\qquad\qquad \textbf{Virtual Lib}\\
   \qquad Sort Real Lib and Virtual Lib by probability in ascending and \\ \qquad descending order\\
   \quad \textbf{Step 3: Create Bi-RNN model\_2}\\
   \qquad 1. Create a causal correlation matrix.\\
   \qquad 2. Construct attack chains based on the matrix.\\
   \qquad 3. Encode the attack chain using Bi-RNN.\\
   \quad \textbf{Step 4: Create classifier on "feint chain" model\_3}\\
   \qquad 1. Divide training sets and test sets(8:2).\\
   \qquad 2. Train and validate model.\\
   \qquad 3. Test model.\\
   \hline
\end{tabular}}
\end{table}

\section{Experimental and Results}
In this section, we discuss the experiment results and give a comprehensive evaluation of bidirectional RNN-based few-shot training for detecting
multi-stage attack model proposed in this paper.
\subsection{Experimental Setup}
\quad\textbf{\emph{A. Experimental Environment}}

We choose same hardware and software configurations when carrying out the experiments. Our experiment is conducted on the operating system of windows 10 on the hardware environment Intel(R) Core(TM) i7-7500U CPU, 8GB RAM and IT hard disk. We utilize the programming language python 3.5. It can be found the main items of our hardware and software configuration in Table~\ref{tab:config}.
\begin{table}[!htbp]
\caption{Hardware and Software Configuration}
\label{tab:config}
\centering
\setlength{\tabcolsep}{0.8mm}{
\begin{tabular}{ccc}
\hline
\hline
\textbf{No.}& \textbf{\emph{Hardware or software}}& \textbf{\emph{Type}}\\
\midrule  
\textbf{1}& Operating system& Windows 10\\
\textbf{2}& Programming language& Python3.5\\
\textbf{3}& Development environment& JetBrains PyCharm 2018.1.4\\
\textbf{4}& CPU& Inter(R) Core(TM)i7-7500U\\
\textbf{5}& RAM& 8GB\\
\textbf{6}& Disk& IT hard disk\\
\hline
\hline
\end{tabular}}
\end{table}

The network topology of DARPA2000 is shown in the following Figure ~\ref{fig:network}, where the network is divided into \emph{DMZ network} and \emph{Inside network}.

\begin{figure}[h]
  \centering
  \includegraphics[width=\linewidth]{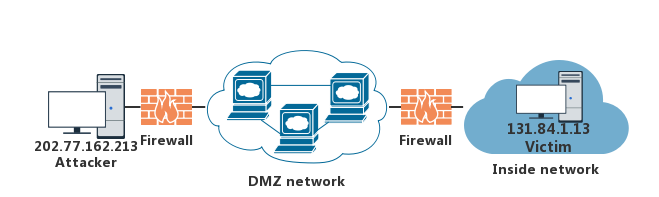}
  \caption{DARPA2000 network topology.}\label{fig:network}
\end{figure}

\textbf{\emph{B. Experimental Data}}

\begin{itemize}
  \item Sadmind Exploit for a DDoS Attack[DARPA2000]:
\end{itemize}

\begin{figure}[h]
  \centering
  \includegraphics[height=7.5cm,width=7cm]{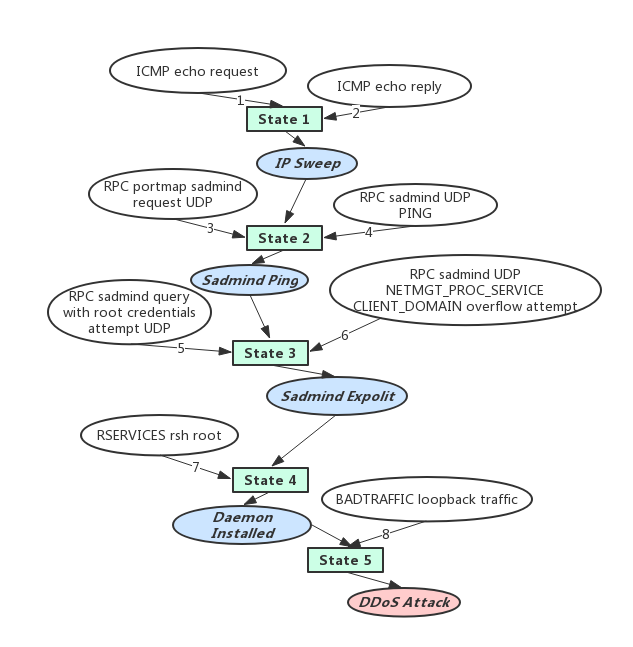}
  \caption{The attack scenario of LLDoS1.0.}\label{fig:process}
\end{figure}

The DARPA2000 dataset is a collection of intrusion scenario correlations from MIT Lincoln lab. It is widely used to verify the effectiveness of various alert event correlation algorithms. LLDOS1.0 includes a complete distributed deny service (DDOS) attack scenario, the multi-stage attack consists of 5 steps. Detect, hack, install trojan mstream DDoS programs and perform remote DDoS attacks on target servers. The attack process is shown in Figure~\ref{fig:process}. The attack mainly utilizes the buffer overflow vulnerability of the sadmind program on the solaris platform. As long as the attacker can correctly find out and overwrite the stack of the executing sadmind program, the attacker can successfully invade the host and obtain the manager from the remote. Permission to execute arbitrary program code, including installing DDoS software and launching DDoS attacks.

\begin{itemize}
  \item Intrusion Detection Evaluation Dataset [CICIDS2017]:
\end{itemize}

The Canadian Institute for Cybersecurity published the CICIDS2017 dataset in 2017. The advantage of this data: time is near, the benchmark data set covers the 11 criteria required, and all previous IDS data sets cannot cover all 11 standards. Containing benign traffic and the latest common attacks, the data capture period begins at 9 am on Monday, July 3, 2017 and ends at 5 pm on Friday, July 7, 2017 for a total of 5 days. Monday is a normal day and only includes benign traffic. Attacks implemented include brute force FTP, brute force SSH, DoS, Heartbleed, web attack, infiltration, botnet and DDoS. They are executed on Tuesday, Wednesday, Thursday and Friday.

\textbf{\emph{C. Evaluation Criteria}}

There are many evaluation indicators used in intrusion detection systems. Although this paper only studies the multi-stage attack identification, it also uses the commonly used indicators in the intrusion detection field, namely the completeness rate and accuracy. Suppose the total number of attacks included in the test data set is $N$, the number of attacks identified by the recognition method is $RN$, and the number of attacks identified in these test data sets is actually $R$. The definitions of these indicators are as follows:

\emph{1) Completeness Rate:} The completeness rate is the completeness of the description method, that is, whether all attacks can be found. The calculation method for multi-stage attack recognition completeness rate is:

\begin{equation}\label{Completeness}
Completeness\ Rate=R/N
\end{equation}

\emph{2) Accuracy Rate:} Accuracy rate is the correctness of the description method, that is, how many of the identified attacks are correct. The calculation method for multi-stage attack recognition accuracy is:

\begin{equation}\label{Accuracy}
Accuracy\ Rate=R/RN
\end{equation}
\subsection{Experimental Result and Evaluation}

\quad \textbf{\emph{A. Alert Correlation Based on Fuzzy Clustering}}

\begin{figure}[!htbp]
  \centering
  \includegraphics[width=\linewidth]{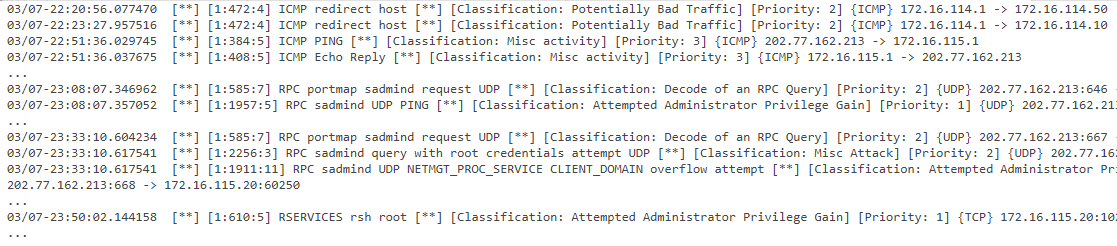}
  \caption{The \emph{Raw Alert} of Snort.}\label{fig:alert}
\end{figure}

Use the snort's command \textbf{\emph{sudo snort -r /LLS\_DDOS\_1.0-inside.\\dump -l /home -A fast -c /etc/snort/snort.conf}} in Linux to replay the original traffic packets from LLDoS1.0 and LLDoS2.0 of DARPA2000 and CICIDS2017. Then, we got the raw alerts of snort in Figure ~\ref{fig:alert}.

Through the network traffic packet analysis software \emph{Wireshark}, we analyzed all traffic packets (including normal background traffic) in the DMZ and Inside areas of LLoS1.0, and the packets containing only attack traffic in each of the five attack phases of the DDoS attack.
\begin{figure}[!htbp]
  \centering
  \includegraphics[height=7.5cm,width=6.5cm]{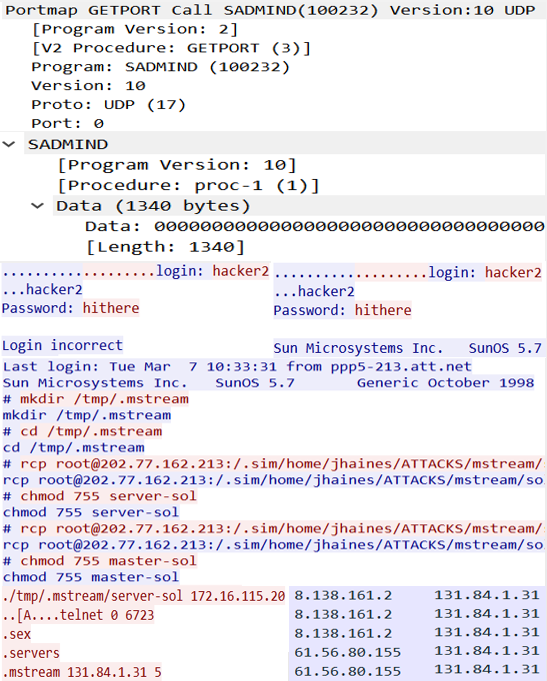}
  \caption{The specific content of the attack traffic packet in wireshark.}\label{fig:wireshark}
\end{figure}
The analysis results are shown in Figure ~\ref{fig:wireshark}. We tracked the TCP flow of the key attack steps and saw that the attacker performed a large number of \emph{IP sweep} (ICMP echo request) on the target network segment, among which 18 hosts survived (ICMP echo reply). The next step is \emph{Sadmind ping}, querying the Sadmind vulnerability and verifying that the service is running on the surviving host. There are 6 hosts that meet this condition. Buffer overflow attacks on these 6 hosts invaded the host, and 3 hosts successfully invaded, namely: 172.16.115.20, 172.16.112.10 and 172.16.112.50. Log in to these three hosts using the rsh service telnet, upload and install the DDoS Daemon (including mstream server and mstream master). Among them, the attacker installed server and master on 172.16.115.20, and only installed server on 172.16.112.10 and 172.16.\\112.50. It can be seen that 172.16.115.20 is the jump host of the attacker in the internal network. Finally, log in to 172.16.115.20, check the port mstream daemon port 6723, execute the mstream command, set the target IP to 131.84.1.31, and use the forged IP to initiate the DDoS attack for 5s.

Combine the two-part alerts (DMZ: 7024 and Inside: 10145) obtained by using snort, and perform alert aggregation on 17169 raw alerts to obtain 3222 alerts. The alert aggregation rate reaches 81.23\%. The result is shown in Table~\ref{tab:freq}.

\begin{table}[!htbp]
  \caption{Alert Aggregation}
  \label{tab:freq}
  \begin{tabular}{|c|c|c|}
    \hline
    Raw Alerts&Amount&Aggregation Rate(\%)\\
    \hline
    DMZ& 7024& --\\
    Inside& 10145& --\\
    Total& 17169& --\\
    Alert Aggregation& 3222& 81.23\\
    \hline
\end{tabular}
\end{table}

Using the fuzzy clustering algorithm proposed in Section 3.1, 3222 alerts are clustered, and a total of 944 attack sequences are obtained. It contains a large number of sequences of length 1 (indicating that there are a large number of fragmentation alerts in the alert clustering).

\begin{table}[!htbp]
\caption{Attack type in $Cluster\ A_2$}
\label{tab:Cluster}
\centering
\setlength{\tabcolsep}{0.8mm}{
\begin{tabular}{p{30pt}<{\centering}p{200pt}}
\hline
\hline
\textbf{No.}& \textbf{Attack type}\\
\midrule  
\textbf{1}& ICMP PING\\
\textbf{2}& FTP Bad login\\
\textbf{3}& TELNET Bad Login\\
\textbf{4}& RPC sadmind UDP PING\\
\textbf{5}& RPC sadmind query with root credentials attempt UDP\\
\textbf{6}& RPC sadmind UDP NETMGT\_PROC\_SERVICE CLIENT\_DOMAIN overflow attempt\\
\textbf{7}& RSERVICES rsh root\\
\textbf{8}& SNMP request udp\\
\textbf{9}& BAD-TRAFFIC loopback traffic\\
\hline
\hline
\end{tabular}}
\end{table}

\begin{figure}[!htbp]
  \centering
  \includegraphics[height=5cm,width=6.5cm]{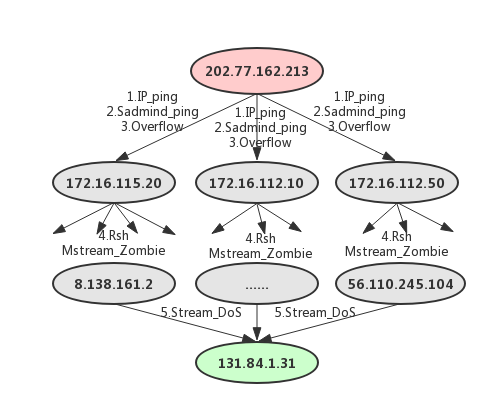}
  \caption{The LLDoS attack process analyzed by Wireshark.}\label{fig:LLDoS}
\end{figure}

After deleting the sequence of length 1, a total of 195 multi-stage attack sequences are obtained. After extracting the multi-stage attack mode, nine sequence patterns are obtained. Among them, the alerts including the multi-stage attack process in LLDOS1.0 are shown in the following Table~\ref{tab:Cluster}. It can be concluded that the attack route is composed of three independent paths in Figure ~\ref{fig:LLDoS}.

\textbf{\emph{B. Building the Virtual-Real Lib}}

By using the downsampling and SMOTE algorithms, the number of our data sets becomes as shown in Table~\ref{tab:pre-processed}.\\

\begin{table*}[!htbp]
\scriptsize
\caption{Balanced dataset}
\label{tab:pre-processed}
\centering
\setlength{\tabcolsep}{0.0000001mm}{
\begin{tabular}{cccccccccccccccc}
\hline
\hline
\multirow{2}*{\textbf{Title of Dataset}} & \multicolumn{15}{c}{\textbf{Data Classified}}\\
\cmidrule(lr){2-16}
\ &\textbf{\emph{Benign}}&\textbf{\emph{DoSHulk}}&\textbf{\emph{PortScan}}&\textbf{\emph{DDoS}}&\textbf{\emph{DoSGoldenEye}}&\textbf{\emph{FTP-Patator}}&\textbf{\emph{SSH-Patator}}&\textbf{\emph{DoSSlowLoris}}&\textbf{\emph{DoSSlowHTTP Test}}&\textbf{\emph{Bot}}&\textbf{\emph{BruteForce}}&\textbf{\emph{XSS}}&\textbf{\emph{Infiltration}}&\textbf{\emph{SQLInjection}}&\textbf{\emph{Heartbleed}}\\
\ \textbf{\emph{Train Dataset}}&1886428&184858&127144&33468&8234&6350&4717&4636&4399&1572&1205&521&28&16&8\\
\ \textbf{\emph{Pre-processed}}&17965&12323&8476&6693&8234&6350&4717&4636&4399&1572&1205&521&280&160&80\\
\hline
\hline
\end{tabular}}
\end{table*}

\begin{figure}[!htbp]
  \centering
  \includegraphics[width=\linewidth]{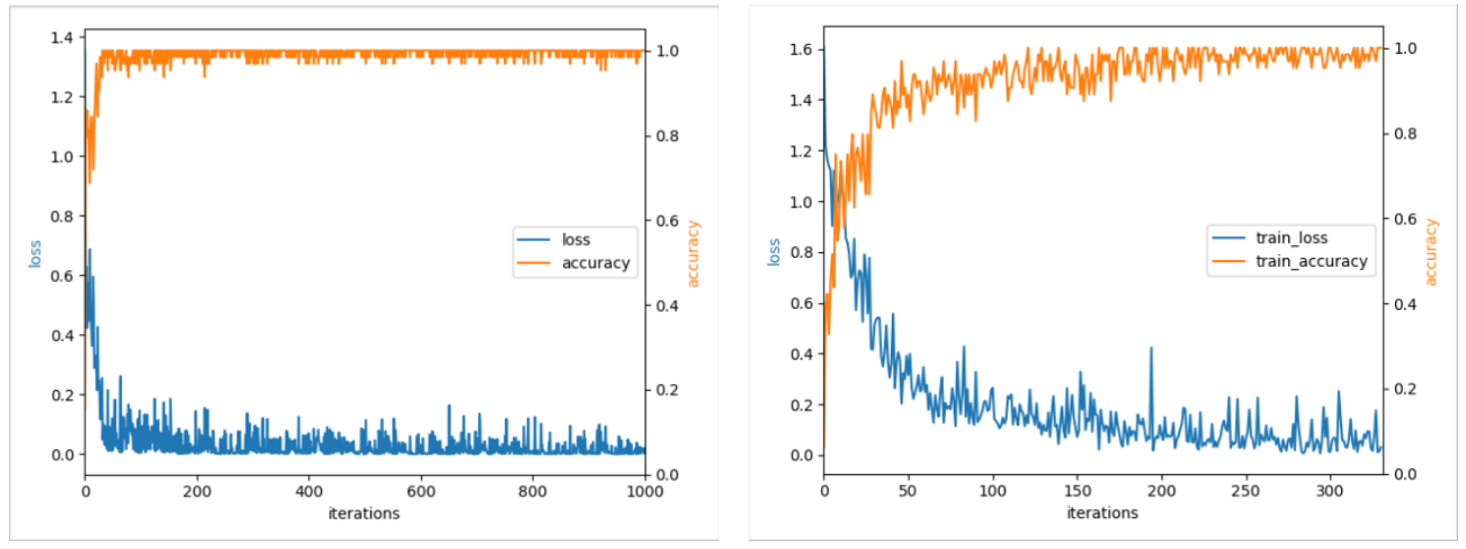}
  \caption{The iteration of the model.}\label{fig:iteration}
\end{figure}

The model using only CNN and using few-shot deep learning model are shown in the Figure ~\ref{fig:iteration}. It can be seen that CNN is easy to cause over-fitting, and the model of few-shot deep learning is used to avoid over-fitting.

The result of few-shot deep learning model is shown in Table ~\ref{tab:result}. We can see that our method has significantly improved the detection rate of Minority class-sample (U2R and R2L).

\begin{table}[h]
\caption{The result of few-shot deep learning model}
\label{tab:result}
\centering
\setlength{\tabcolsep}{0.00000005mm}{
\begin{tabular}{|c|c|c|c|c|c|c|c|}
\hline
\multicolumn{2}{|c|}{ \multirow{2}*{\textbf{Confusion Matrix}} }& \multicolumn{5}{c|}{Predicted Category} &\multirow{2}*{\textbf{\emph{Recall}}}\\
\cline{3-7}
\multicolumn{2}{|c|}{}&Benign&Probe&DoS&U2R&R2L&\\
\hline
\multirow{5}*{Actual}&Benign&60352&123&103&9&6&0.996\\
\cline{2-8}
&Probe&387&3501&260&0&18&0.840\\
\cline{2-8}
&DoS&5686&82&224081&0&4&0.975\\
\cline{2-8}
&U2R&73&13&17&119&6&0.522\\
\cline{2-8}
&R2L&7018&4&6&1&9160&0.566\\
\cline{1-8}
\multicolumn{2}{|c|}{\textbf{\emph{Precision}}}&0.821&0.940&0.998&0.922&0.996&\textbf{\emph{Acc:0.96}}\\
\hline
\end{tabular}}
\end{table}

\begin{figure}[!htbp]
  \centering
  \includegraphics[height=4.4cm,width=6.5cm]{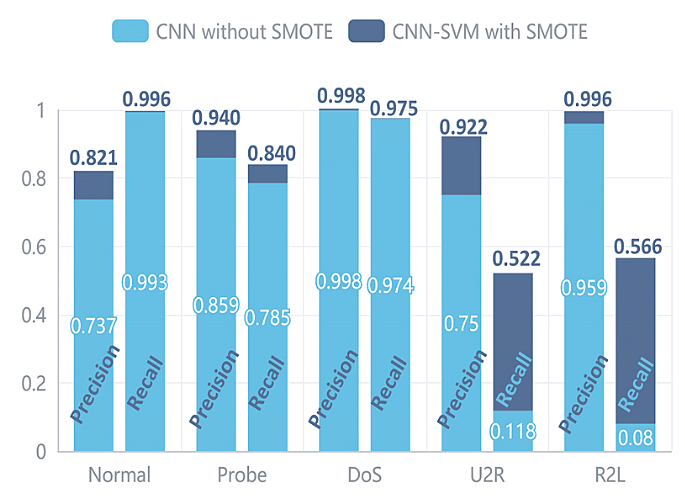}
  \caption{Comparison of precision and recall.}\label{fig:PB}
\end{figure}

We find the fact that CNN-SVM with SMOTE get the best recall and precision. CNN model without SMOTE has lower recall when the
classify U2R and R2L traffic. The reason of that the amount of U2R and R2L packages is too lower than other packages what we have mentioned above. But the recall to U2R and R2L traffic has been greatly improved by introduced SMOTE. The result can be seen in Figure ~\ref{fig:PB}.

\begin{figure}[!htbp]
  \centering
  \includegraphics[height=4.5cm,width=6.5cm]{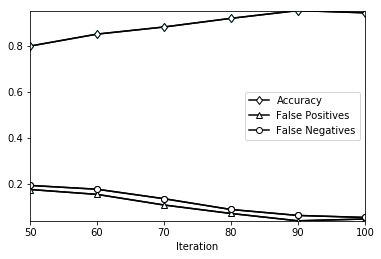}
  \caption{The effect of the number of iterations on the experimental results.}\label{fig:i}
\end{figure}

The evaluation criteria of the number of iterations is shown in Figure ~\ref{fig:i}.

The Virtual-Real Lib contains 20,718 real attacks and 189,826 virtual attacks. In order to verify the reliability of our results, we got all of real attacks and \emph{Normal} to test, the results show that more than 99\% of the real attacks are missed as normal.

\textbf{\emph{C. Build Feint Lib and Detect the "Feint Attack"}}

\begin{figure}[!htbp]
  \centering
  \includegraphics[height=3.5cm,width=8cm]{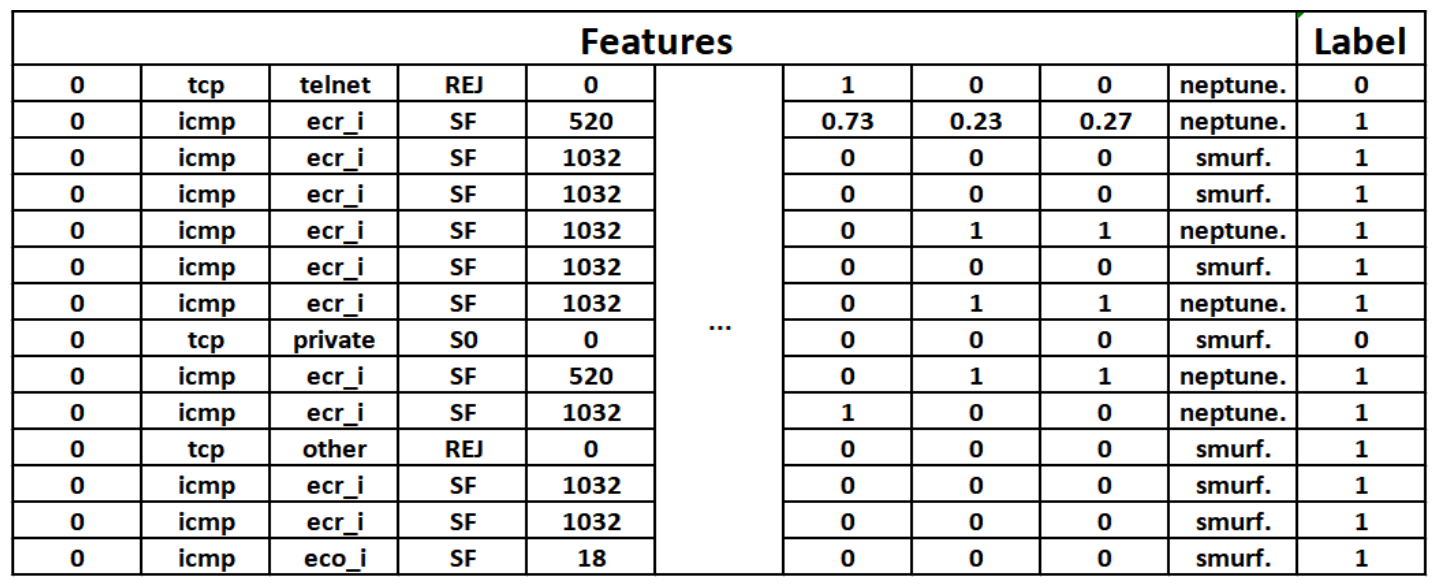}
  \caption{The Feint Lib.}\label{fig:feint}
\end{figure}

Feint Lib contains 11758 records of "Feint Attack" chains, and there has 20 attacks in each record. The number of training sets is 9408 and the number of testing sets is 2350. The dataset can be seen in Figure ~\ref{fig:feint}. Label 1 means the chain is a "Feint Attack" chain. Label 0 means the chain is a common chain.

The number of real attacks in the attack chain are 1 to 7. Among them, the number of attack chains containing one real attack is 3371, the number of attack chains containing two real attacks is 3248, the number of attack chains containing three real attacks is 1811, the number of attack chains containing four real attacks is 672, the number of attack chains containing five real attacks is 200, the number of attack chains containing six real attacks is 50, the number of attack chains containing seven real attacks is 11, and the number of attack chains containing eight real attacks is one.

\begin{figure}[!htbp]
  \centering
  \includegraphics[height=4.5cm,width=6cm]{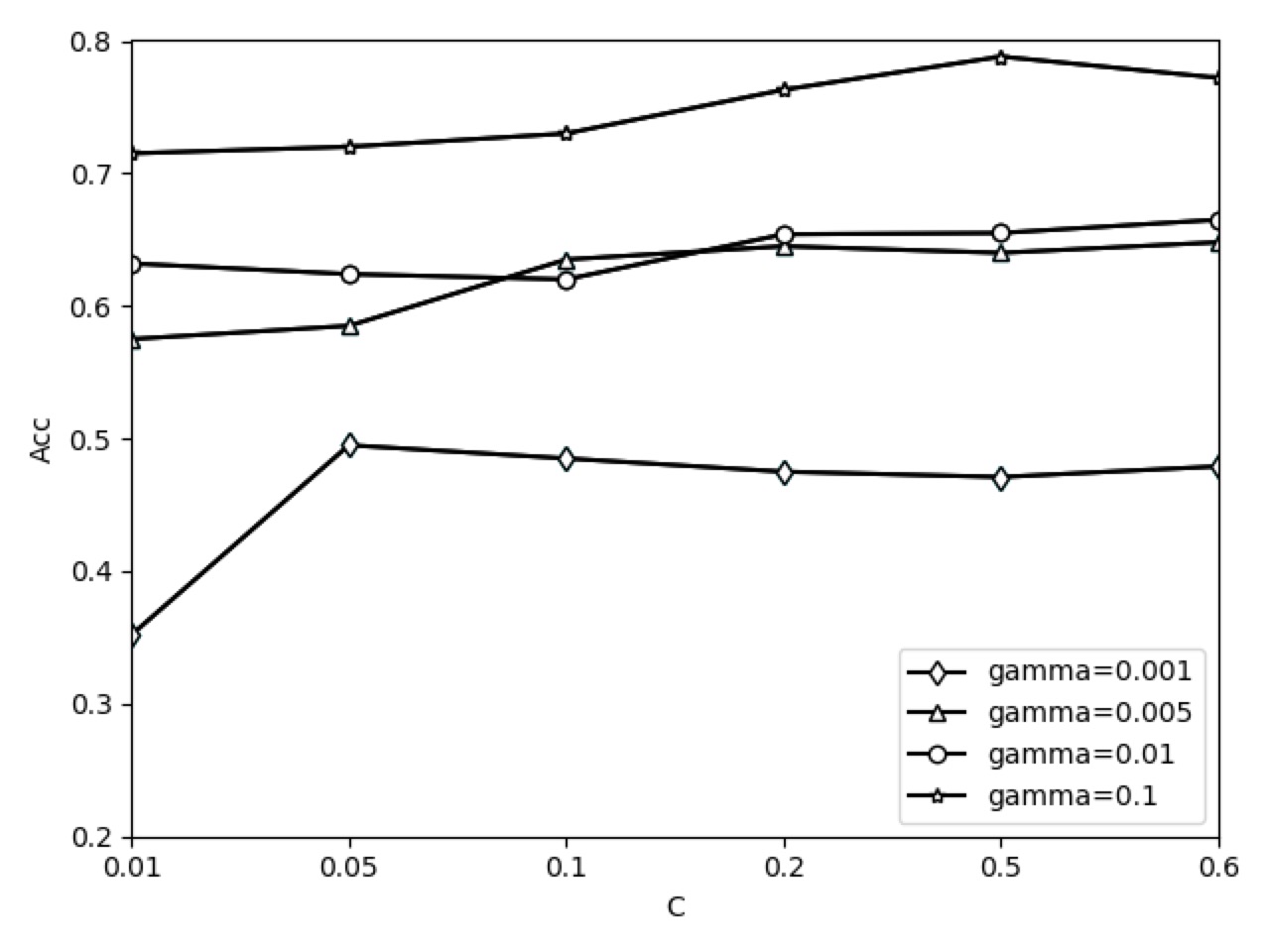}
  \caption{The result of detecting "Feint Attack" chain.}\label{fig:svm}
\end{figure}

Finally, we chose $best\ c$ = 0.5 and $best\ g$=1 to get the $best\ acc$ = 78.8764\% of cross validation. We got 75.23\% accuracy on the test set. It is shown in Figure ~\ref{fig:svm}.
\section{Conclusion}

In this paper, aiming at the "Feint Attack" mode in APT attack, we proposed new detection method which mainly utilizes fuzzy clustering and Bi-RNN algorithm. Firstly, by analyzing the existing "Feint Attack", we defined virtual attacks and real attacks as the basic attack events that constitute the "Feint Attack" chain. In the attack scenario, the fuzzy clustering method based on attribute similarity is used to mine multi-stage attack chains. A multi-stage attack mode comparison library is formed, and a few-shot deep learning model is defined and divided into virtual attacks and real attacks to construct a dataset of atomic attack events. Then, the atomic attack event is embedded into the attack chain through Bi-RNN coding, and the "Feint Attack" chain is constructed to form the "Feint Attack" dataset. Finally, the attack chain samples containing the feint attack behavior and the non-feint attack behavior are further classified to achieve the purpose of accurately identifying the "Feint Attack". Our innovation lies in the first use of bidirectional RNN coding to construct the attack chain to ensure maximum retention of causal information. We verified our method by using the LLDoS1.0 and LLDoS2.0 of DARPA2000 and CICIDS2017 of Canadian Institute for Cybersecurity. The experimental results show that our method can derive the multi-stage attack sequence from the alert correlation by fuzzy clustering, and the "Feint Attack" behavior is mined from the attack chains. The attack sequence is encoded by Bi-RNN, and achieve 75.23\% accuracy to identify "Feint Attack".Research on the key technologies of behavior detection, and realize the prototype system based on the virtual attack and real attack chain to achieve zero breakthrough in detecting such attacks.

%
\begin{acks}
This research is supported by the Fundamental Research Funds for the Central Universities of China under Grants (No.2018JBZ103), Science and Technology on Information Assurance Laboratory (No.\\614200103011711), the National Natural Science Foundation of China (No.61672092), Beijing Excellent Talent Training Project (No.B-MK2017B02-2), the Fundamental Research Funds for the Central Universities (No.2017RC016), and China Scholarship Council (CSC No.201807095014).
\end{acks}

%
\bibliographystyle{ACM-Reference-Format}
\bibliography{sample-base}
%
\appendix

\end{document}